\documentclass[%
 aip,
 jcp,
 amsmath,amssymb,
 reprint,%
]{revtex4-1}

\usepackage[english]{babel}
\usepackage{graphicx}
\usepackage{dcolumn}
\usepackage{bm}
\usepackage[version=3]{mhchem}
\usepackage{multirow}

\usepackage[utf8]{inputenc}
\usepackage[T1]{fontenc}
\usepackage{mathptmx}
\usepackage{xcolor}

\begin{document}

\preprint{AIP/123-QED}

\title{Competition of quantum effects in H$_2$/D$_2$ sieving in carbon
 nanotubes}

\author{Manel Mondelo-Martell}
\email{manel.mondelo@chemie.uni-frankfurt.de}
\altaffiliation[Current address:]
{Institut f\"ur Physikalische und Theoretische Chemie, J.W. Goethe Universit\"at
Frankfurt}
\author{Ferm\'in Huarte-Larra\~naga}
\affiliation
{Department of Materials Science \& Physical Chemistry and Institute of Theoretical and Computational Chemistry (IQTCUB), Universitat de Barcelona, Barcelona}

\date{\today}

\begin{abstract}
  Nanoporous materials have the potential to be used as \emph{molecular sieves} to separate chemical substances in a mixture via selective adsorption and kinetic sieving. The separation of isotopologues is also possible via the so-called \emph{quantum sieving} effect: the different effective size of isotopologues due to their different Zero Point Energy (ZPE). Here we compare the diffusion rates of Hydrogen and Deuterium in (8,0)~Single Walled Carbon Nanotubes obtained with quantum dynamics simulations. The diffusion channels obtained present important contributions from resonances connecting the potential wells. These resonances, which are more important for \ce{H2} than for \ce{D2}, increase the low-temperature diffusivity of both isotopologues, but prevent the inverse kinetic isotope effect reported for similar nanostructured systems.

\end{abstract}

\maketitle

\section{Introduction}
The importance of hydrogen in chemical research and industry cannot be overestimated: from its use as reagent in synthetic chemistry and several industrial processes, to its potential application as a clean fuel for combustion batteries, the future of our society seems to be inevitably tied to being able to harness hydrogen's capabilities as efficiently as possible\cite{Dunn2002,Schlapbach2001,Basile2017}. One particular aspect of such harnessing is the separation of hydrogen and deuterium, since the latter has very different applications in the areas of isotopological tracing\cite{Stiopkin2011}, proton nuclear magnetic resonance spectroscopy\cite{Mantsch1977,Ewanicki2020}, neutron scattering\cite{Buldt1978, Liebschner2018}. If chemical separation processes are extremely costly \emph{per se}, using around 10-15\% of the total amount of energy consumed worldwide\cite{AboutSeparations}, isotopological separation is comparatively the most expensive\cite{HandbookNucChem, Spindel1991}. In the particular case of \ce{H2}/\ce{D2} separation, at industrial level the most common technique is the cryogenic distillation, which achieves a separation factor of merely 1.5\cite{RAE1978}. Finding a more efficient pathway to achieve this separation is clearly one of the main objectives in current research in the area of chemical separation\cite{Basile2017}. A very promising alternative  is based on the \emph{quantum sieving} effect, proposed by Beenakker \emph{et al}\cite{Beenakker1994}. This phenomenon is a consequence the mass difference between \ce{H2} and \ce{D2}: a lower mass increases the zero-point energy of the internuclear bond, which in turn results in a more diffuse wave function and therefore a larger \emph{effective size}. Thus, \ce{H2} has a larger effective size when compared with the heavier \ce{D2}. This size difference becomes critical when the molecules enter nanometric cavities, and can affect both their adsorption and diffusion properties. In the last two decades, the adsorption of \ce{H2} and \ce{D2} has been studied in several nanoporous materials such as carbon nanotubes\cite{Wang1999,Lu2006,Garberoglio2009, Rather2020, Li2020}, zeolites\cite{DeLuca2004, Salazar2019, Radola2020NewZeolites,Bezverkhyy2020}, or metal--organic frameworks (MOFs)\cite{Oh2016,FitzGerald2008,Kim2017,Fitzgerald2018, Cao2020, Wang2020} with the aim of finding the best candidate for isotopic separation of \ce{H2} and \ce{D2}; see also Ref~\citenum{Kim2019} and references therein. Very recently, specifically tailored organic cage molecules have been reported to obtain selectivities of up to 8 by combining small pores connecting large cavities\cite{Liu2019}.

When studying quantum sieving, one has to distinguish between the change of the adsorption and diffusion properties of the adsorbates. In the former case, a heavier isotopologue is preferentially adsorbed in the nanomaterial (\emph{thermodynamic} or \emph{chemical affinity} quantum sieving), which can be straightforwardly interpreted through the relative change in ZPE of the two species when entering the nanometric cavity. Regarding diffusion, it is known that there are two competing quantum effects which simultaneously play a role in the process:
on one hand, the ZPE effects described above decrease the diffusion barrier for the heavier deuterium, resulting in a higher mobility than expected. This effect has been claimed to result in an \emph{inverse kinetic isotope effect}, namely, a faster diffusion of deuterium in nanoporous materials, compared with hydrogen\cite{Kumar2005, Kumar2006, Kumar2008}. However, these studies were based on semiclassical Transition State Theory simulations, which neglect the second quantum effect: resonant tunneling, which has been seen to enhance hydrogen transport properties in systems such as carbon nanotubes\cite{Mondelo-Martell2017}.
In this work we revisit the case of H2/D2 quantum sieving in single walled
carbon nanotubes and provide accurate diffusion rates for for both molecules in
the low pressure regime. The results reported here evidence that by
significantly extending the propagation time up to 20 picoseconds we are able to
resolve quantum resonances below the diffusion barrier that change drastically
the perspective of the previously published simulations. These much larger
propagation times, required by the low corrugation of the potential energy
profile along the nanotube axis, are achieved thanks to an adiabatic approach
described in a previous work\cite{Mondelo-Martell2017}.

\section{Theoretical Methods}
\subsection{Diffusion coefficient calculation}\label{subs:diffcoeff}

Previous theoretical works\cite{Kumar2006, Kumar2008, Hankel2011, Mondelo-Martell2016, Mondelo-Martell2017} on the diffusion of hydrogen in
nanoporous carbon, regardless of the specific potential energy surface
employed, show that the interaction between the molecule and the nanostructure
generates a potential energy profile that consists of collection of minima
(adsorption sites) separated by maxima (diffusion barriers) along the nanotube
axis. Following previous studies\cite{Zhang1999} and given the shape of the potential
energy, we have modelled the  molecular diffusion process a set of uncorrelated
jumps between neighbouring adsorptions sites\cite{Doll1987, Barth2000}. From
this perspective, the diffusion rate is obtained through:
\begin{equation}
	D_{\mathit{diff}}=\frac{l^2}{2d} k_{\mathit{hop}}(T). \label{eq:Ddiff}
\end{equation}
where $k_{\mathit{hop}}$ is the hopping probability between adjacent sites, and $d$ is the dimensionality of the system (1 in this case).
The problem of calculating the diffusion coefficient is then reduced to the calculation of $k_{\mathit{hop}}$. Following Zhang and Light\cite{Zhang1999}, this probability will be calculated through the flux--correlation function approach in a quantum dynamics formalism, which is summarized below.

The general expression for a transition rate is given by the thermal average of the Cumulative Reaction Probability (CRP), $N(E)$, which gives us the probability that the system has to go from any reactant state to any product state, as a function of the energy. Then, assuming a Boltzmann distribution for the energy of the system, we have:
\begin{equation}
	k(T)=\frac{1}{2 \pi Q(T)}\int_{-\infty} ^{\infty} e^{- \beta E}N(E) dE.\label{eq:k}
\end{equation}
In the previous equation, $\beta=\frac{1}{k_BT}$, and $Q(T)$ is the partition function of the system. In our case, the partition function of molecular hydrogen or deuterium is factorized as a product of its (5) confined  degrees of freedom and the unconfined diffusion coordinate (z):
\begin{equation}
    Q(T)=Tr(e^{\beta\hat{H}^{5D}}) q_z(T)
\end{equation}
with $q_z(T)=L\left(\frac{mT}{2\pi}\right)^{\frac{1}{2}}$ the semiclassical
partition function of a particle in a periodic potential and $\hat{H}^{5D}$ the
Hamiltonian of the \emph{confined coordinates} of the system (see below).
The origin of energy was chosen to be the minimum value of the PES (note that the
origin of energy of both the partition function and $N(E)$ has to be chosen
consistently to ensure that $k(T)$ does not depend on the energy reference
taken). Finally, we use the flux--correlation functions 
approach\cite{Yamamoto1960, Miller1974,Miller1983} to compute $N(E)$, as
implemented in Refs~\citenum{Matzkies1998,Manthe2008a}. In this approach one
sets a dividing surface, $h$, which separates reactants from products and 
defines the thermal flux operator accross such a surface as 
$\hat{F}_{T_0}=e^{-\beta_0 \hat{H}/2} i [\hat{H},h]e^{-\beta_0 \hat{H}/2}$, with
$\beta=1/k_BT_0$. The eigenvalues and eigenstates of this operator ($f_{T_0}$
and $|f_{T_0}\rangle$, respectively) are obtained by iterative diagonalization
and then propagated in time to compute a flux correlation function,
$C_{ff}(t)=\sum_{m,n}f_mf_n\langle f_n\mid e^{- \mathrm{i}\hat{H}t}\mid f_m \rangle$.
The CRP is finally obtained by Fourier transform of $C_{ff}(t)$:
\begin{align} \label{eq:NEflux2}
	N(E)&=\frac{1}{2}e^{2\beta_0 E}\sum_{n} \sum _{m}f_n f_m \Biggl \lvert  \int_{-\infty}^{\infty} dt e^{\mathrm{i} Et}\langle f_n\mid e^{- \mathrm{i}\hat{H}t}\mid f_m \rangle \Biggr \rvert ^2.
\end{align}
In this work we have used the State-Averaged variant of the Multiconfigurational Time--dependent Hartree (SA-MCTDH)\cite{Manthe2008a} for both the iterative diagonalization of the different operators, and the propagation of the resulting wavepackets.

\subsection{Modelization of the system}
In order to compute the transition rate in Eq.~\ref{eq:k}, we have modelled a single \ce{H2} (or \ce{D2}) molecule in the hollow cavity of a carbon nanotube using its 3 internal degrees of freedom ($\rho$, $\theta$, $\phi$) as well as the 3 translational DOFs of the molecular center of mass ($x$, $y$, $z$) as a coordinate system. This representation has been previously used by ourselves \cite{Mondelo-Martell2015b, Mondelo-Martell2016, Mondelo-Martell2017} as well as by other authors\cite{Lu2006, Garberoglio2006}. The quantum dynamics calculations have been carried out using an adiabatic approximation, as described in a previous publication\cite{Mondelo-Martell2017}. Following such model we have solved the dynamics of a 1D wavepacket on a set of particular potential energy surfaces instead of propagating a full 6D function. The approximation is justified in terms of a time--scale separation argument between the large--amplitude motion of the molecule along the axis of the nanotube ($z$) and the fast motion on the \emph{confined} coordinates (vibration, rotation, and translation in the $xy$ plane; collectively referred to as $q$). The molecular wave packet evolves, thus, according to the 1D Time--dependent Schr\"odinger Equation (note that atomic units are used throughout this paper, and therefore $\hbar=1$):
\begin{equation}
i \frac{\partial}{\partial t} \tilde{\psi}_j(z,t) = \hat{H}_{j}^{(ad)} \tilde{\psi}_j(z,t),\label{eq:ad-TDSE}
\end{equation}
with the adiabatic Hamiltonian defined as:
\begin{equation}
\hat{H}_{j}^{(ad)} = \frac{1}{2m}\frac{\partial^2}{\partial z^2} z + \sum _{k=1}^{N_z} \varepsilon_j(z_k) 
\left|z_k \middle\rangle \middle\langle z_k  \right| \label{eq:Hadiab}
\end{equation}
Here, each function $\varepsilon_j(z_k)$ represents the $z$--dependent eigenvalue of a given eigenstate $\xi_j^{5D}(q;z_k)$ of the confined coordinates Hamiltonian, $\hat{H}^{5D}(q,z_k)$, and acts effectively as a potential energy term for the motion of the wave packet along the $z$ dimension. We will hence refer to each $\varepsilon_j(z_k)$ as a \emph{confined eigenstate potential energy surface} (cePES). To obtain them, we diagonalize the Hamiltonian of the confined coordinates at different points along the $z$ axis; each $z$--dependent eigenvalue then conforms a given cePES. 

In a previous work we proved that this adiabatic representation yields excellent agreement with full-dimensional propagations of the same system, while drastically reducing the computational effort. For details on the derivation and the model of the 5D eigenstates, we refer the reader to Ref.~\citenum{Mondelo-Martell2017}.

\section{Results and discussion}\label{sec:results}
\subsection{Confined Eigenstates Potential Energy Surfaces}\label{subsec:cePES}
As a first step to study the diffusion, we took advantage of the adiabatic approach to visualize the diffusion process as a 1D problem. By observing the computed cePES we can obtain relevant information on the diffusion mechanism even before running actual quantum dynamics simulations of that process. 

The SA-MCTDH approach was used to compute the eigenstates of the 5D Hamiltonian at the center of a unit cell of the nanotube ($z=0$) via iterative diagonalization, using the same wave function representation parameters reported in \cite{Mondelo-Martell2017} and listed in Table~\ref{tab:5Deig} for the sake of clarity.
\begin{table}
	\caption{Primitive and SPF basis sets for the calculation of the eigenstates of the 5D Hamiltonian for both H$_2$ and D$_2$. Distances given in Bohr radii, angles in radians. The same primitive basis was used for both isotopologues.}\label{tab:5Deig}
	\begin{tabular}{c|cc|ccc}\hline
		\multirow{2}{*}{DOF}&\multicolumn{2}{c}{Number of SPFs} \vline& \multicolumn{3}{c}{Primitive grid}\\
        					&H$_2$&D$_2$	&Num. Points&Type&Range \\ \hline
        $\rho$				&	2		&	2			&	32	&	FFT	&	0.5--5.0\\
        $\theta$			&	5		&	5			&	64	&cot-DVR&	0--$\pi/2$\\
        $\phi$				&	7		&	7			&	64	&	FFT	&	0--$2\pi$\\
        $x$					&	3		&	4			&	32	&	FFT	&	-3.5--3.5\\
        $y$					&	3		&	4			&	32	&	FFT	&	-3.5--3.5\\ \hline
	\end{tabular}
\end{table}
The 50 lowest energy eigenstates, $\xi^{5D}_j(q;z)$, were used as a basis for the matrix representation of $\hat{H}^{5D}$ at 512 equispaced points along the $z$ coordinate, extending from $-56.1$ to $56.1$ Bohr radii which corresponds to 14 unit cells of the SWCNT. The total number of carbon atoms used to define the interaction potential was large enough to ensure that no edge effects were noticeable at the ends of the simulation grid.

\begin{figure*}
\hspace{-0.75cm}
\begin{minipage}[c]{0.5\linewidth}
\centering
\includegraphics[scale=0.65]{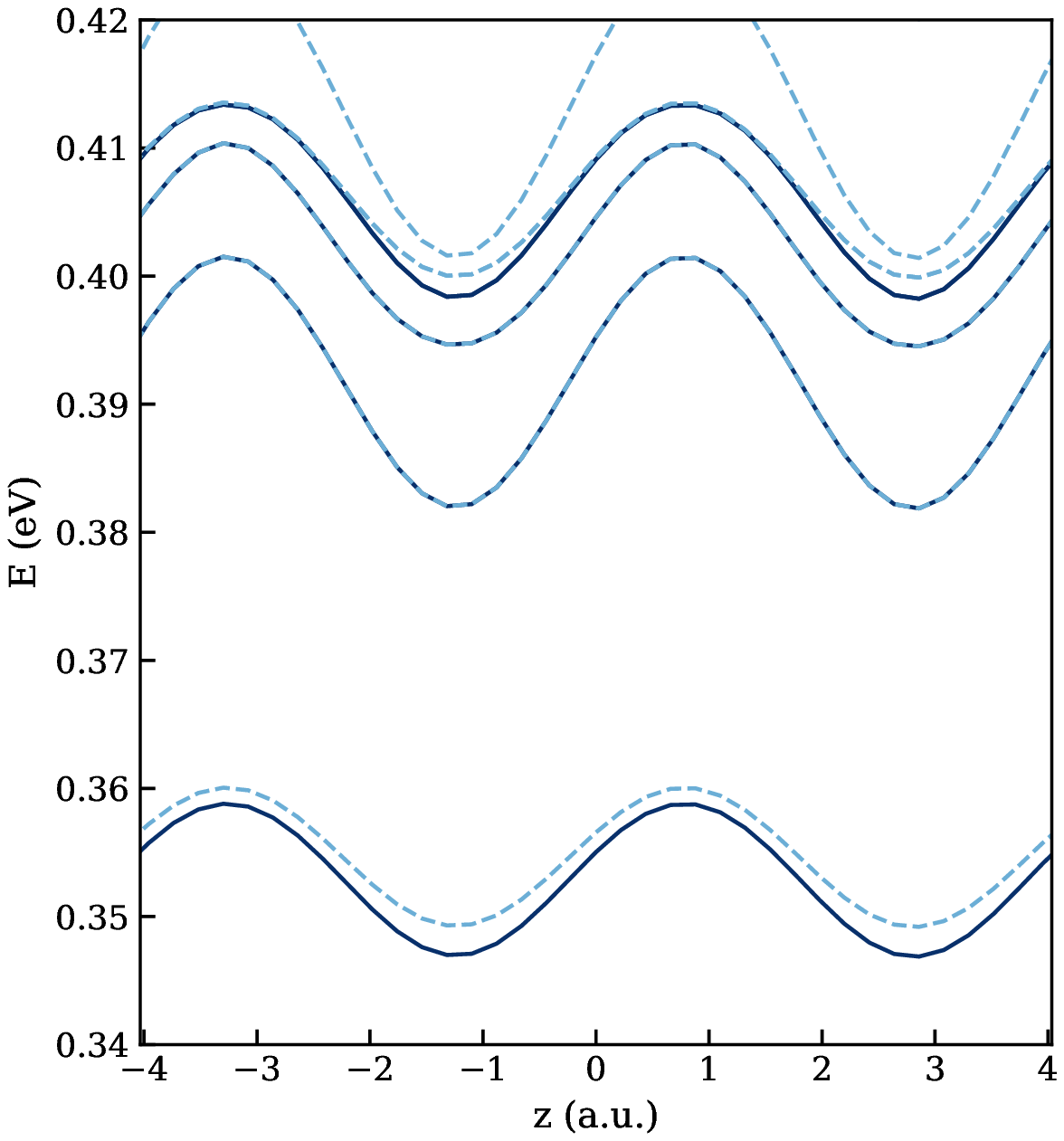}
\end{minipage}
\begin{minipage}[c]{0.5\linewidth}
\centering
\includegraphics[scale=0.65]{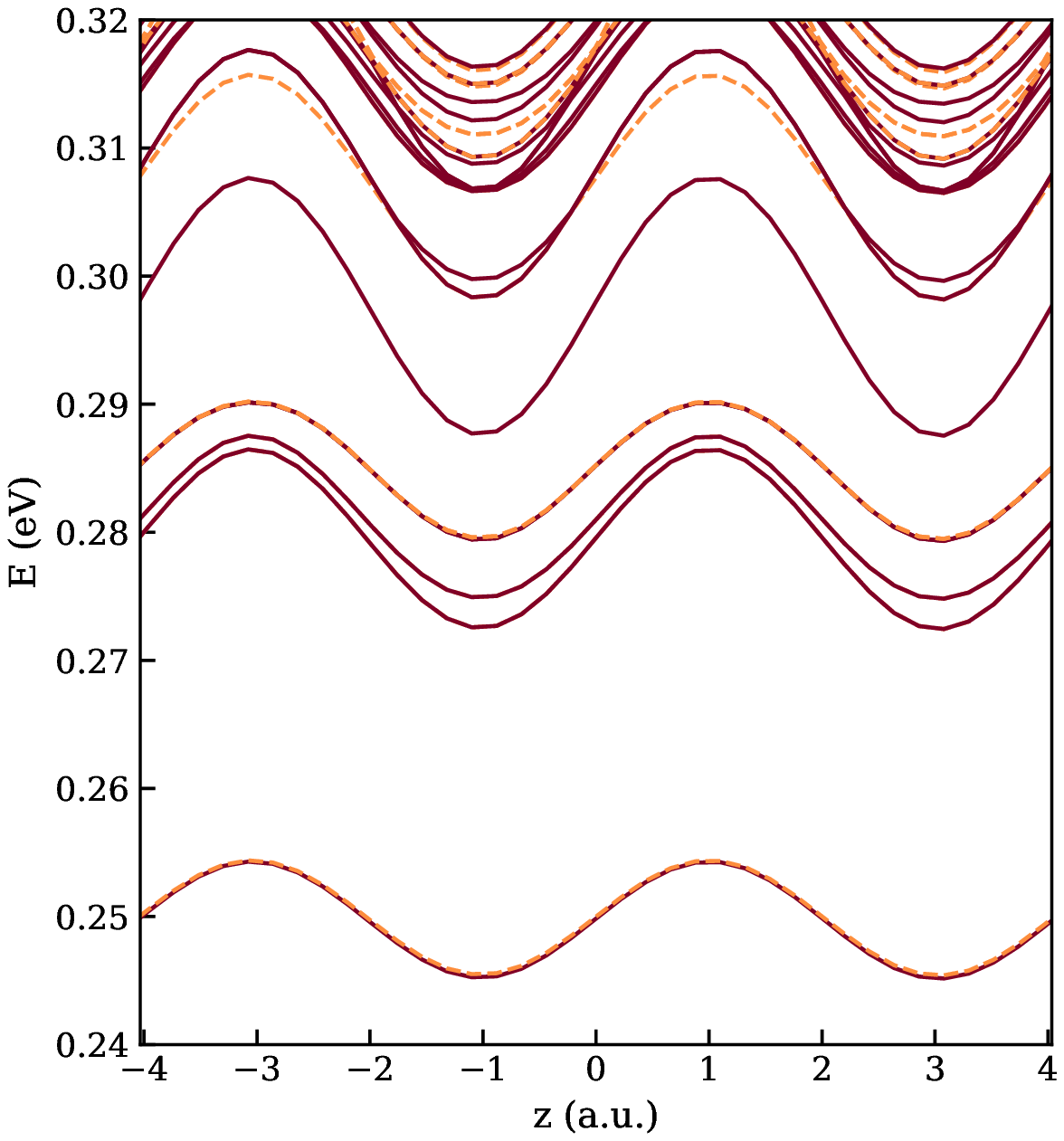}
\end{minipage}
\caption{Variation along a single unit cell of the eigenenergies of the 5D hydrogen (left) and deuterium (right) eigenstates (cePES). Symmetric eigenstates with respect to inversion are represented with solid lines, while dashed lines label asymmetric states.}\label{fig:cePES}
\end{figure*}

The lower energy cePES corresponding to the \ce{H2} molecule are depicted in the left panel of Figure~\ref{fig:cePES}. The different curves are drawn according to the \emph{ortho}--\emph{para} symmetry of the 5D eigenstate they represent: the solid, darker lines label a symmetric state and the dashed, lighter lines label an antisymmetric state. The same scheme is followed in the right panel, which shows the curves for \ce{D2}.
From these figures we can extract an approximate value of the threshold energy, which is the minimum energy (including ZPE) the particle would need to go from reactants to products. This quantity corresponds to the maximum of the lowest--energy cePES, and has a value of $E_{\mathit{tr}}=0.3588$~eV for \ce{H2}, while for \ce{D2} it decreases to $E_{\mathit{tr}}=0.2543$~eV. The adiabatic diffusion barrier, $\Delta E_{\mathit{tr}}$, defined as the difference between the maximum and the minimum of each cePES, is also slightly lower for \ce{D2} (5.6~meV) than for \ce{H2} (7.2~meV). A final feature is the density of confined eigenstates that each molecule presents. We can use the relation between flux eigenstates and the vibrational states of the transition complex to estimate how many eigenstates will contribute to the diffusion for a given energy range from the Boltzmann population of the vibrational levels of the molecule fixed at a point along the diffusion coordinate. It is readily seen that the spectrum is denser for \ce{D2} than for \ce{H2}, as a consequence of its larger mass, and therefore more eigenstates will contribute to the diffusion for \ce{D2} than for \ce{H2}.

All the effects presented in this section are directly or indirectly related to the different ZPE of the isotopologues, and point in the direction that, in confined environments, \ce{D2} could diffuse faster than the lighter \ce{H2}. This \emph{inverse kinetic isotope effect} has been described previously in Carbon Molecular Sieves\cite{Kumar2005, Kumar2008, Nguyen2010, Hankel2011, Contescu2013a} and nanotubes\cite{Mondelo-Martell2016}, and justified as a purely ZPE effect: due to the lower ZPE of \ce{D2} its effective size is smaller than that of \ce{H2}, and as a consequence the heavier isotopologue feels less the corrugation of the potential generated by the Carbon atoms, thus diffusing more easily than \ce{H2}. In the next section, we will investigate if this effect persists when including all possible quantum effects via time--dependent quantum dynamics simulation of the system.

\subsection{Hopping probabilities and Diffusion Coefficients}
As outlined in Section~\ref{subs:diffcoeff}, we used the flux correlation functions approach together with the SA-MCTDH method to compute the probability  associated to the molecule transitioning between adjacent adsoprtion sites. Since quantum effects are more relevant at lower temperatures, we set a value of $\beta$ corresponding to 100~K, which yields numerically stable results for $N(E)$ in the range of 40 to 125~K. The qualitative analysis of the cePES above allowed us to estimate that only 6 to 8 states have an significant population at the maximum temperature considered, and therefore contribute appreciably to $k(T)$. To ensure convergence of the calculations we have computed a total of 26~flux eigenstates to represent the diffusion process of the \ce{H2} molecule, and 30 in case of \ce{D2}. The basis set used to represent the MCTDH wave function in both cases is described in Table~\ref{tab:flux}. 
\begin{table}
	\caption{Primitive and SPF representation used in the flux eigenstates calculation and propagation.}\label{tab:flux}
	\begin{tabular}{c|cc|ccc}\hline
		\multirow{2}{*}{DOF}&\multicolumn{2}{c}{Number of SPFs} \vline& \multicolumn{3}{c}{Primitive grid}\\
        					& H$_2$ & D$_2$ &Num. Points&Type&Range \\ \hline
        $q$					&	20	&	14	&	50	&	Discrete	&	--\\
        $z$					&	20	&	20	&	512	&	FFT	&	-56.066\ --\ 56.066~$a_0$\\\hline
	\end{tabular}
\end{table}
 
The resulting flux states were propagated for a total time of 20~ps, using again the SA-MCTDH method.
To prevent transmissions and reflections in the edge of the representation grid we added a transmission--free complex absorbing potential (CAP) as defined by \citeauthor{Manolopoulos2002}\cite{Manolopoulos2002,Gonzalez-Lezana2004} with a length of 20~bohr. The transmission--free nature of the CAP was required due to the existence of long--lived processes in the diffusion mechanism (See below, and Ref.\citenum{Mondelo-Martell2017}, for details on these processes).
 
\begin{figure*}
\hspace{-0.75cm}
\begin{minipage}[c]{0.5\linewidth}
\centering
\includegraphics[scale=0.55]{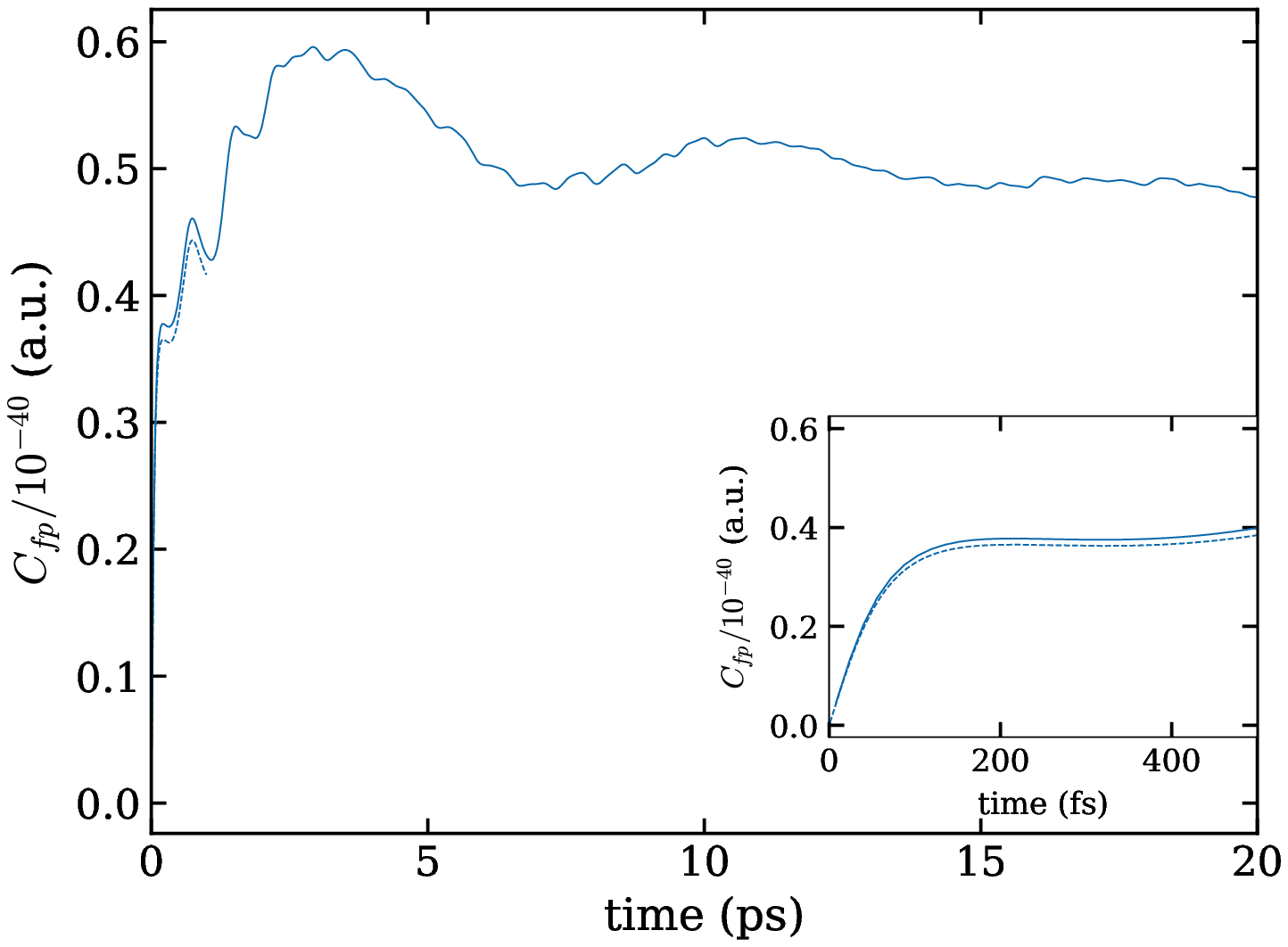}
\end{minipage}
\begin{minipage}[c]{0.5\linewidth}
\centering
\includegraphics[scale=0.55]{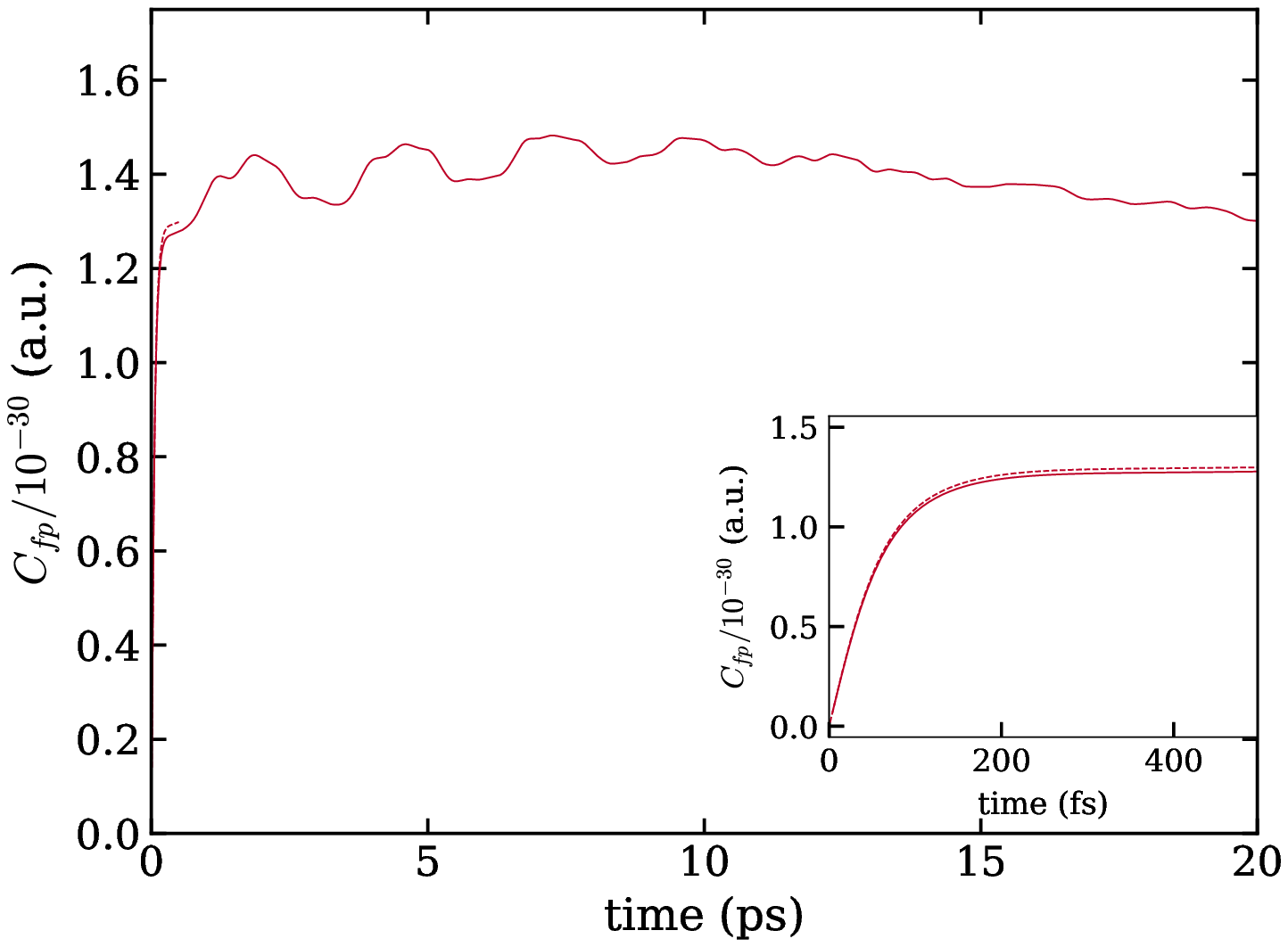}
\end{minipage}
\caption{Flux--position correlation function for the diffusion of \ce{H2} (right) and \ce{D2} (left) along an (8,0)~SWCNT. Inset: close up to the short--time region (dashed line corresponding to the 6D propagation in Ref.~\citenum{Mondelo-Martell2016}).}\label{fig:Cfp}
\end{figure*}

The integral of $C_{ff}$ over time $t$, known as the \emph{flux--position correlation function}, $C_{fp}(t)$, can be used as a rule of thumb to estimate the convergence of a calculation: the value of this quantity stabilizes as the wave packet leaves the interaction region, and it reaches a plateau once the amplitude of the function in that region becomes zero. Therefore, to obtain a perfectly converged CRP one has to choose a propagation time $T$ such that a constant value of $C_{fp}(t)$ is achieved for any time $t>T$. The flux--position correlation functions obtained after a total propagation time of 20~ps for \ce{H2} and \ce{D2} are shown in the left and right panel of Figure~\ref{fig:Cfp}, respectively.

For the long time propagations reported in this work, note that in neither case
does the function admittedly reach a constant value, thus indicating that a
portion of the wave packets is still in the interaction region, even after
20~ps. This is a direct consequence of the small barrier for the diffusion
process. Additionally, the fine structure of the functions is also a signal of
the presence of resonances in the diffusion process\cite{Mondelo-Martell2017}.
These two features of the system would enforce us to go to much longer times to
obtain a perfectly converged CRP. However, this time would only contribute to
the resolution of the fine structure of the resonances, at the price of much 
more computational effort and potential numerical instability. Instead of this,
we decided to fix the maximum energy resolution for the calculation of the CRP
by multiplying $C_{ff}(t)$ by a Gaussian convolution function with 
$\Delta E=0.12$~meV. This reduces the aliasing coming from the truncation of the
Fourier Transform in Eq.~\ref{eq:NEflux2}\cite{Mondelo-Martell2017}, yielding
a smoother function with better convergence properties and, as long as the
value of $\Delta E$ is smaller than $k_B T$ for all the temperature range
studied, not causing significant errors to the calculation of the transition
rate, Eq.~\eqref{eq:k}. To confirm that the CRP are sufficiently converged at
20~ps, we compared the resulting functions after an increase of 10\% on the total
propagation time, getting essentially the same results for both $N(E)$ and the
transition rate.

\begin{figure}
\centering
\includegraphics[scale=0.6]{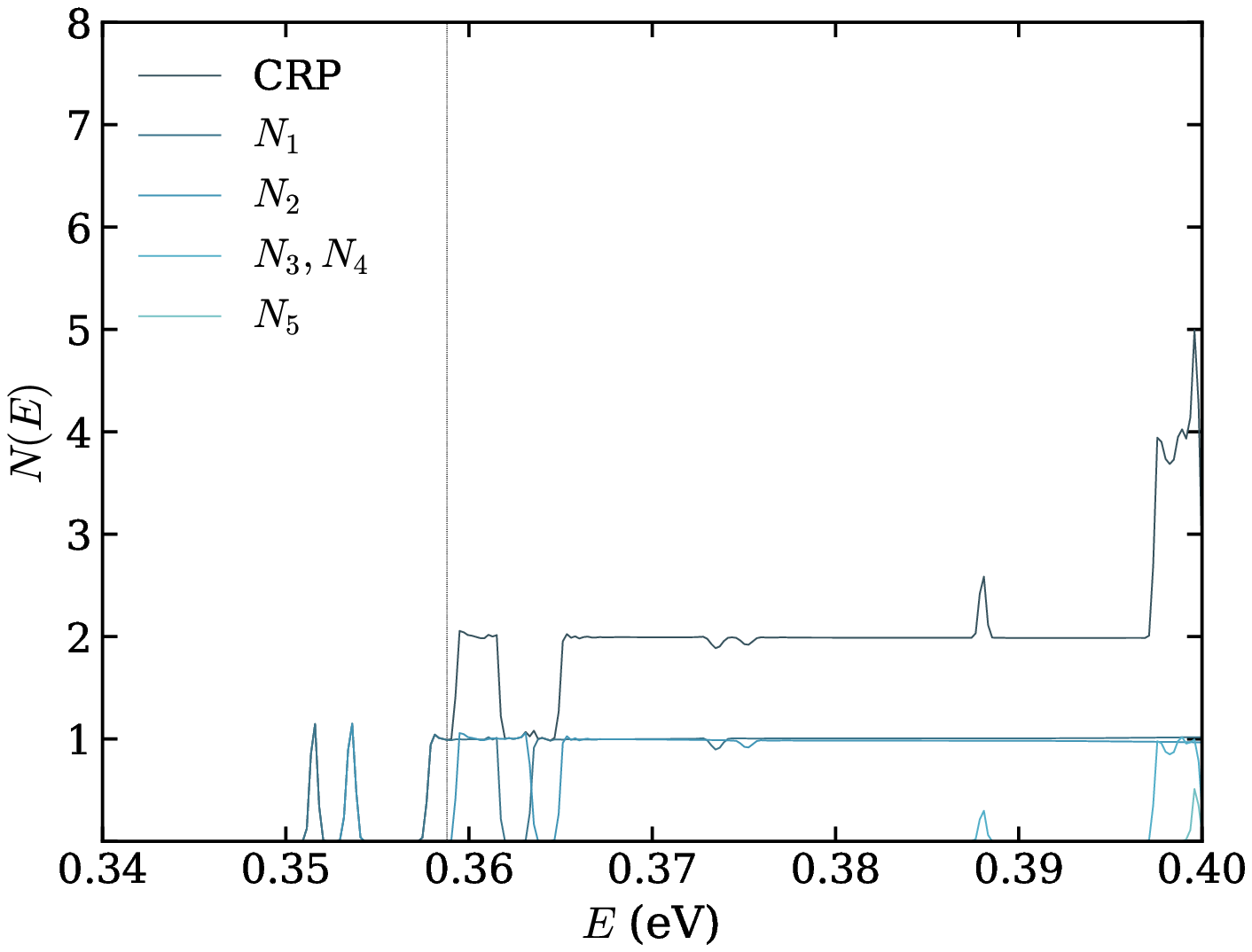}
\caption{Cumulative Reaction Probability and individual flux eigenstates contributions for the diffusion of \ce{H2} after 20~ps propagation. Vertical dotted lines marks the diffusion energy threshold.}\label{fig:NEH2}
\end{figure}

\begin{figure}
\centering
\includegraphics[scale=0.6]{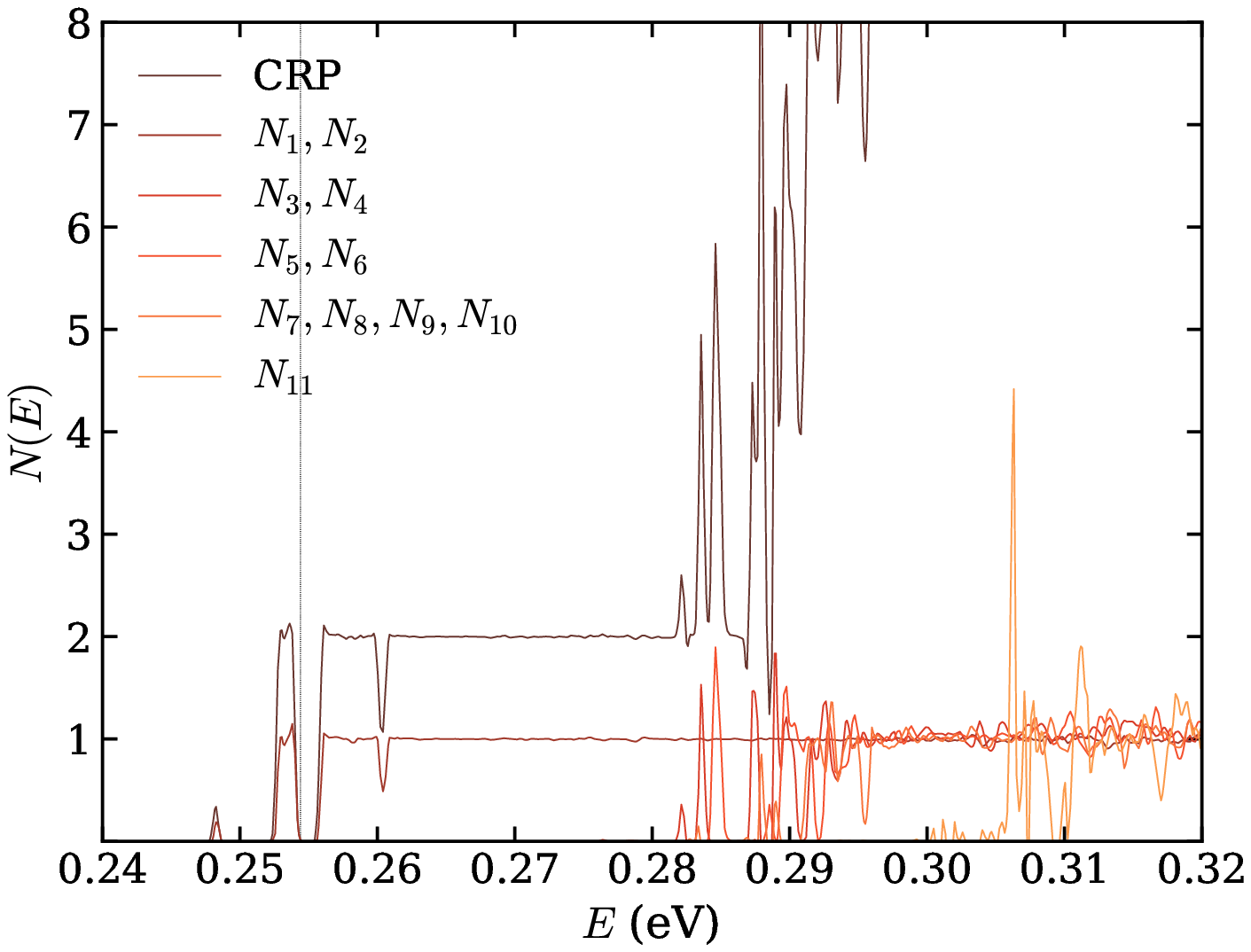}
\caption{Cumulative Reaction Probability and individual flux eigenstates contributions for the diffusion of \ce{D2} after 20~ps propagation. Vertical dotted lines marks the diffusion energy threshold.}\label{fig:NED2}
\end{figure}

The resulting $N(E)$ for \ce{H2} and \ce{D2} are shown in Figures~\ref{fig:NEH2} and~\ref{fig:NED2}, respectively, together with the individual contributions of the lowest--energy flux eigenstates. Additionally, the threshold energy for the diffusion process is marked in both figures as a dashed vertical line.

The first feature to notice in both plots is their significant amount of fine
structure in form of sharp peaks. In a previous work\cite{Mondelo-Martell2017}
these features were confirmed to be shape and Feshbach resonances by calculating
the exact 6D~eigenstates of \ce{H2} in a unit cell of the carbon nanotube using
periodic boundary conditions, and checking that the energies of the peaks
coincide with the eigenvalues of the tunneling 6D states. Moreover, the width of
the peaks is consistent with the tunneling splitting of the resonant state,
which further confirms the accuracy of the propagation. The large amount of
resonant states is consistent with the shape of $C_{fp}(t)$ discussed
previously, as the low-energy resonances provide a way for the wave packet to
enter and leave the interaction region easily. More importantly, some resonances
are found at energies below the diffusion threshold for both isotopologues,
indicating that tunneling effect is relevant for the diffusion process at low
temperatures. 

Comparing the cumulative reaction probability curves for both \ce{H2} and
\ce{D2}, one can see two main differences: on one hand, the energy and intensity
of the first resonances; on the other, the density of the higher energy
resonances. The first feature is probably the more critical point, since this
difference will influence more heavily the behavior of the different molecules
at very low temperature. The two first resonances in $N(E)$ for the \ce{H2}
molecule are both intense and appear at energies significantly lower than the
threshold. Moreover, a second-order resonance appears at $E=0.36$~eV, just above
the diffusion threshold, thus contributing to increase the diffusion rate at all
temperatures. On the contrary, for \ce{D2} we have a very weak (and therefore 
negligible) resonance at $E=0.248$~eV, while two strong resonances exist at
energies just below the energy threshold. However, since they are so close to
the diffusion threshold, these resonances will have a smaller effect on
$D_{\mathit{diff}}$ than those present in \ce{H2}. This can be understood by
noticing that, in the calculation of the transition probability,
Eq.~\eqref{eq:k}, the CRP has to be weighted by a Boltzmann distribution and
normalized with the partition function of the system. The latter avoids
dependencies on the absolute energy scale chosen as reference, so that only
relative differences in energy matter.
Thus, as a result of the exponentially decaying Boltzmann function, the relative
weight of the below-barrier resonances becomes larger in the overall diffusion
probability as the temperature diminishes. This effect becomes more noticeable as
the energy difference between the tunneling resonances and the diffusion barrier
increases. In the case of \ce{H2} diffusion, resonances are so apart in energy that,
at very low temperature values, the two below-barrier resonances dominate the
overall diffusion probability. Conversely, for deuterium the resonances are close enough to the barrier to 
contribute almost the same as the first above-barrier diffusion states at any
given temperature value. This also explains why the effects of low--energy resonances decreases
as temperature increases: as the above-barrier region gains weight due to the
Boltzmann function, more diffusion states get populated, so the relative weight of the individual
resonant states decreases. Similarly, even though \ce{D2} presents a denser resonance
spectrum than \ce{H2} at high energies, these do not play any role in the
diffusion process in the studied temperature range, since the resonant states
are too energetic to have significant population compared with the set of regular
diffusion states.

\begin{figure}
	\includegraphics[scale=0.6]{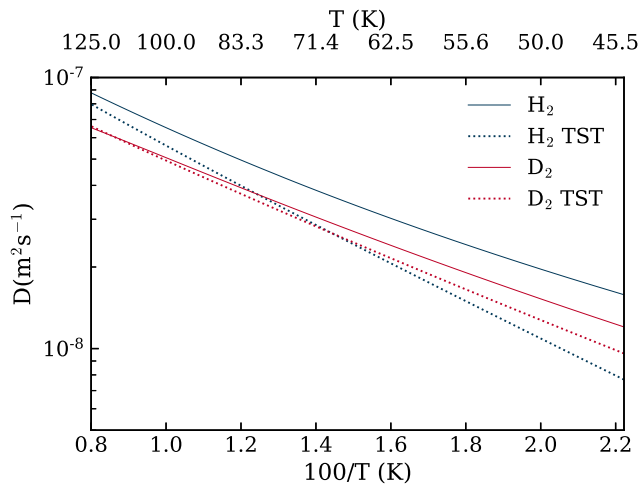}
	\caption{Diffusion rates for \ce{H2} (blue) and \ce{D2} (red) computed with the flux correlation function approach (solid lines) and with TST (dotted lines). }\label{fig:Drate}
\end{figure}

Once we analyzed $N(E)$ thoroughly, the transition coefficient has been obtained by Boltzmann averaging of the CRP at different temperatures, and then inserted in Eq~\eqref{eq:Ddiff} to compute the diffusion rate. The diffusion coefficient is plotted in Figure~\ref{fig:Drate} as a function of temperature inverse for both \ce{H2} and \ce{D2} as solid blue and red lines, respectively. The same quantities have been computed using a Transition State Theory (TST) model, and are also shown in the same figure as dotted lines.
The sum of all the quantum effects outlined previously (\emph{i.e.} ZPE effects and tunneling) significantly changes the diffusion rates with respect to those predicted by simple TST calculations. It is readily seen that there is a relevant increase of $D_{\mathit{diff}}$ at low temperatures for both isotopologues, but more so for the lighter \ce{H2}, as it was expected from the discussion of the cumulative reaction probability above. In fact, some tunneling contributions remains at temperatures as high as 125~K for this molecule. The consequence of the different amount of tunneling effect for both species is important even from a qualitative point of view: TST predicts a turnover of the diffusion rates at temperatures below 70~K, with \ce{D2} starting to diffuse faster than \ce{H2} at this point. This inverse Kinetic Isotope Effect is consistent with the discussion of the adiabatic diffusion barrier discussed Section~\ref{subsec:cePES} and widely studied in Carbon Molecular Sieves. In a previous work\cite{Mondelo-Martell2016} we reported that the same effect would be observed for the diffusion of \ce{H2} and \ce{D2} along SWCNTs, and supported such a claim with quantum dynamics simulations of the full 6D system up to 500~fs. Instead, having been able to extend our quantum simulations to a remarkable limit such as 20 ps has changed dramatically our conclusions: 500 fs propagations provide insufficiently converged hopping probability, unable to resolve the sharp below tunneling resonances. With the correctly converged cumulative reaction probability and the resonances properly resolved, tunneling outweighs the ZPE effects at low temperatures, and the inverse kinetic isotope effect does not take place. These results show the importance of an accurate quantum mechanical description of \ce{H2} and \ce{D2} when studying their diffusion properties.

\section{Summary and Conclusions}
The calculation of the diffusion rates for \ce{H2} and \ce{D2} along a (8,0)~CNT in the low pressure limit has been carried out using the single--hopping approach. The hopping rate was obtained through the general expression of a transmission rate, with the cumulative reaction probability computed with the flux--correlation function approach. In order to achieve convergence of $N(E)$, we used an adiabatization scheme to reduce the problem from a 6D Hamiltonian to a 2D system, thus being able to propagate the flux eigenstates for 20~ps. This allowed us to resolve resonant structures in $N(E)$ which enhance diffusion at low temperatures.

The diffusion rates calculated in this work differ from previous studies in that no inverse kinetic isotope effect appears in this particular system. This qualitative inconsistency with a previous work of us indicates that $N(E)$ was not correctly converged in those calculations. The difference with other theoretical and experimental studies on other nanostructured materials like Carbon Molecular Sieves, however, is probably due to the different structure of those, which present large pores connected through narrow channels rather than the cylindrical shape of carbon nanotubes. Despite this fundamental difference, the behavior of the molecules inside a carbon nanotube can be used to model the transport within the narrow channels connecting pores in other nanomaterials, or to consider the design of new devices based purely on nanotubes.

\section{Acknowledgements}
Financial support from the Spanish Ministerio de Econom\'ia y Competitividad (Ministry of Economy and Competitiveness) (CTQ2013-41307-P) and Generalitat de Catalunya (2014-SGR-25) is acknowledged. M.M.-M. further thanks a predoctoral grant from the FPU program (FPU2013/02210) from the Spanish Ministerio de Educaci\'on, Cultura y Deporte (Ministry of Education, Culture and Sports). 


\begin{thebibliography}{50}%
\makeatletter
\providecommand \@ifxundefined [1]{%
 \@ifx{#1\undefined}
}%
\providecommand \@ifnum [1]{%
 \ifnum #1\expandafter \@firstoftwo
 \else \expandafter \@secondoftwo
 \fi
}%
\providecommand \@ifx [1]{%
 \ifx #1\expandafter \@firstoftwo
 \else \expandafter \@secondoftwo
 \fi
}%
\providecommand \natexlab [1]{#1}%
\providecommand \enquote  [1]{``#1''}%
\providecommand \bibnamefont  [1]{#1}%
\providecommand \bibfnamefont [1]{#1}%
\providecommand \citenamefont [1]{#1}%
\providecommand \href@noop [0]{\@secondoftwo}%
\providecommand \href [0]{\begingroup \@sanitize@url \@href}%
\providecommand \@href[1]{\@@startlink{#1}\@@href}%
\providecommand \@@href[1]{\endgroup#1\@@endlink}%
\providecommand \@sanitize@url [0]{\catcode `\\12\catcode `\$12\catcode
  `\&12\catcode `\#12\catcode `\^12\catcode `\_12\catcode `\%12\relax}%
\providecommand \@@startlink[1]{}%
\providecommand \@@endlink[0]{}%
\providecommand \url  [0]{\begingroup\@sanitize@url \@url }%
\providecommand \@url [1]{\endgroup\@href {#1}{\urlprefix }}%
\providecommand \urlprefix  [0]{URL }%
\providecommand \Eprint [0]{\href }%
\providecommand \doibase [0]{http://dx.doi.org/}%
\providecommand \selectlanguage [0]{\@gobble}%
\providecommand \bibinfo  [0]{\@secondoftwo}%
\providecommand \bibfield  [0]{\@secondoftwo}%
\providecommand \translation [1]{[#1]}%
\providecommand \BibitemOpen [0]{}%
\providecommand \bibitemStop [0]{}%
\providecommand \bibitemNoStop [0]{.\EOS\space}%
\providecommand \EOS [0]{\spacefactor3000\relax}%
\providecommand \BibitemShut  [1]{\csname bibitem#1\endcsname}%
\let\auto@bib@innerbib\@empty
\bibitem [{\citenamefont {Dunn}(2002)}]{Dunn2002}%
  \BibitemOpen
  \bibfield  {author} {\bibinfo {author} {\bibfnamefont {S.}~\bibnamefont
  {Dunn}},\ }\href {\doibase 10.1016/S0360-3199(01)00131-8} {\bibfield
  {journal} {\bibinfo  {journal} {International Journal of Hydrogen Energy}\
  }\textbf {\bibinfo {volume} {27}},\ \bibinfo {pages} {235} (\bibinfo {year}
  {2002})}\BibitemShut {NoStop}%
\bibitem [{\citenamefont {Schlapbach}\ and\ \citenamefont
  {Z{\"{u}}ttel}(2001)}]{Schlapbach2001}%
  \BibitemOpen
  \bibfield  {author} {\bibinfo {author} {\bibfnamefont {L.}~\bibnamefont
  {Schlapbach}}\ and\ \bibinfo {author} {\bibfnamefont {A.}~\bibnamefont
  {Z{\"{u}}ttel}},\ }\href {\doibase 10.1038/35104634} {\bibfield  {journal}
  {\bibinfo  {journal} {Nature}\ }\textbf {\bibinfo {volume} {414}},\ \bibinfo
  {pages} {353} (\bibinfo {year} {2001})}\BibitemShut {NoStop}%
\bibitem [{\citenamefont {Basile}\ \emph {et~al.}(2017)\citenamefont {Basile},
  \citenamefont {Dalena}, \citenamefont {Jianhua~Tong},\ and\ \citenamefont
  {Vezirolu}}]{Basile2017}%
  \BibitemOpen
  \bibinfo {editor} {\bibfnamefont {A.~B.}\ \bibnamefont {Basile}}, \bibinfo
  {editor} {\bibfnamefont {F.~D.}\ \bibnamefont {Dalena}}, \bibinfo {editor}
  {\bibfnamefont {J.~T.}\ \bibnamefont {Jianhua~Tong}}, \ and\ \bibinfo
  {editor} {\bibfnamefont {T.~N.~V.}\ \bibnamefont {Vezirolu}},\ eds.,\ \href
  {\doibase 10.1049/PBPO089E} {\emph {\bibinfo {title} {Hydrogen Production,
  Separation and Purification for Energy}}}\ (\bibinfo  {publisher}
  {Institution of Engineering and Technology},\ \bibinfo {year}
  {2017})\BibitemShut {NoStop}%
\bibitem [{\citenamefont {Stiopkin}\ \emph {et~al.}(2011)\citenamefont
  {Stiopkin}, \citenamefont {Weeraman}, \citenamefont {Pieniazek},
  \citenamefont {Shalhout}, \citenamefont {Skinner},\ and\ \citenamefont
  {Benderskii}}]{Stiopkin2011}%
  \BibitemOpen
  \bibfield  {author} {\bibinfo {author} {\bibfnamefont {I.~V.}\ \bibnamefont
  {Stiopkin}}, \bibinfo {author} {\bibfnamefont {C.}~\bibnamefont {Weeraman}},
  \bibinfo {author} {\bibfnamefont {P.~A.}\ \bibnamefont {Pieniazek}}, \bibinfo
  {author} {\bibfnamefont {F.~Y.}\ \bibnamefont {Shalhout}}, \bibinfo {author}
  {\bibfnamefont {J.~L.}\ \bibnamefont {Skinner}}, \ and\ \bibinfo {author}
  {\bibfnamefont {A.~V.}\ \bibnamefont {Benderskii}},\ }\href {\doibase
  10.1038/nature10173} {\bibfield  {journal} {\bibinfo  {journal} {Nature}\
  }\textbf {\bibinfo {volume} {474}},\ \bibinfo {pages} {192} (\bibinfo {year}
  {2011})}\BibitemShut {NoStop}%
\bibitem [{\citenamefont {Mantsch}, \citenamefont {Sait{\^{o}}},\ and\
  \citenamefont {Smith}(1977)}]{Mantsch1977}%
  \BibitemOpen
  \bibfield  {author} {\bibinfo {author} {\bibfnamefont {H.~H.}\ \bibnamefont
  {Mantsch}}, \bibinfo {author} {\bibfnamefont {H.}~\bibnamefont
  {Sait{\^{o}}}}, \ and\ \bibinfo {author} {\bibfnamefont {I.~C.}\ \bibnamefont
  {Smith}},\ }\href {\doibase 10.1016/0079-6565(77)80010-1} {\bibfield
  {journal} {\bibinfo  {journal} {Progress in Nuclear Magnetic Resonance
  Spectroscopy}\ }\textbf {\bibinfo {volume} {11}},\ \bibinfo {pages} {211}
  (\bibinfo {year} {1977})}\BibitemShut {NoStop}%
\bibitem [{\citenamefont {Ewanicki}, \citenamefont {Kim},\ and\ \citenamefont
  {Wang}(2020)}]{Ewanicki2020}%
  \BibitemOpen
  \bibfield  {author} {\bibinfo {author} {\bibfnamefont {J.}~\bibnamefont
  {Ewanicki}}, \bibinfo {author} {\bibfnamefont {W.}~\bibnamefont {Kim}}, \
  and\ \bibinfo {author} {\bibfnamefont {W.}~\bibnamefont {Wang}},\ }\href
  {\doibase 10.1002/mrc.5027} {\bibfield  {journal} {\bibinfo  {journal}
  {Magnetic Resonance in Chemistry}\ }\textbf {\bibinfo {volume} {58}},\
  \bibinfo {pages} {733} (\bibinfo {year} {2020})}\BibitemShut {NoStop}%
\bibitem [{\citenamefont {B{\"{u}}ldt}\ \emph {et~al.}(1978)\citenamefont
  {B{\"{u}}ldt}, \citenamefont {Gally}, \citenamefont {Seelig}, \citenamefont
  {Seelig},\ and\ \citenamefont {Zaccai}}]{Buldt1978}%
  \BibitemOpen
  \bibfield  {author} {\bibinfo {author} {\bibfnamefont {G.}~\bibnamefont
  {B{\"{u}}ldt}}, \bibinfo {author} {\bibfnamefont {H.~U.}\ \bibnamefont
  {Gally}}, \bibinfo {author} {\bibfnamefont {A.}~\bibnamefont {Seelig}},
  \bibinfo {author} {\bibfnamefont {J.}~\bibnamefont {Seelig}}, \ and\ \bibinfo
  {author} {\bibfnamefont {G.}~\bibnamefont {Zaccai}},\ }\href {\doibase
  10.1038/271182a0} {\bibfield  {journal} {\bibinfo  {journal} {Nature}\
  }\textbf {\bibinfo {volume} {271}},\ \bibinfo {pages} {182} (\bibinfo {year}
  {1978})}\BibitemShut {NoStop}%
\bibitem [{\citenamefont {Liebschner}\ \emph {et~al.}(2018)\citenamefont
  {Liebschner}, \citenamefont {Afonine}, \citenamefont {Moriarty},
  \citenamefont {Langan},\ and\ \citenamefont {Adams}}]{Liebschner2018}%
  \BibitemOpen
  \bibfield  {author} {\bibinfo {author} {\bibfnamefont {D.}~\bibnamefont
  {Liebschner}}, \bibinfo {author} {\bibfnamefont {P.~V.}\ \bibnamefont
  {Afonine}}, \bibinfo {author} {\bibfnamefont {N.~W.}\ \bibnamefont
  {Moriarty}}, \bibinfo {author} {\bibfnamefont {P.}~\bibnamefont {Langan}}, \
  and\ \bibinfo {author} {\bibfnamefont {P.~D.}\ \bibnamefont {Adams}},\ }\href
  {\doibase 10.1107/S2059798318004588} {\bibfield  {journal} {\bibinfo
  {journal} {Acta Crystallographica Section D: Structural Biology}\ }\textbf
  {\bibinfo {volume} {74}},\ \bibinfo {pages} {800} (\bibinfo {year}
  {2018})}\BibitemShut {NoStop}%
\bibitem [{Abo()}]{AboutSeparations}%
  \BibitemOpen
  \href {http://www.cchem.berkeley.edu/co2efrc/about.html} {\enquote {\bibinfo
  {title} {{About | Center for Gas Separations}},}\ }\BibitemShut {NoStop}%
\bibitem [{\citenamefont {Van~Hook}(2003)}]{HandbookNucChem}%
  \BibitemOpen
  \bibfield  {author} {\bibinfo {author} {\bibfnamefont {W.~A.}\ \bibnamefont
  {Van~Hook}},\ }in\ \href {\doibase 10.1007/0-387-30682-X{\_}44} {\emph
  {\bibinfo {booktitle} {Handbook of Nuclear Chemistry}}}\ (\bibinfo
  {publisher} {Springer US},\ \bibinfo {address} {Boston, MA},\ \bibinfo {year}
  {2003})\ pp.\ \bibinfo {pages} {1863--1897}\BibitemShut {NoStop}%
\bibitem [{\citenamefont {Spindel}\ and\ \citenamefont
  {Ishida}(1991)}]{Spindel1991}%
  \BibitemOpen
  \bibfield  {author} {\bibinfo {author} {\bibfnamefont {W.}~\bibnamefont
  {Spindel}}\ and\ \bibinfo {author} {\bibfnamefont {T.}~\bibnamefont
  {Ishida}},\ }\href {\doibase 10.1021/ed068p312} {\bibfield  {journal}
  {\bibinfo  {journal} {Journal of Chemical Education}\ }\textbf {\bibinfo
  {volume} {68}},\ \bibinfo {pages} {312} (\bibinfo {year} {1991})}\BibitemShut
  {NoStop}%
\bibitem [{\citenamefont {Rae}(1978)}]{RAE1978}%
  \BibitemOpen
  \bibfield  {author} {\bibinfo {author} {\bibfnamefont {H.~K.}\ \bibnamefont
  {Rae}},\ }in\ \href {\doibase 10.1021/bk-1978-0068.ch001} {\emph {\bibinfo
  {booktitle} {Separation of Hydrogen Isotopes}}},\ Vol.~\bibinfo {volume}
  {10},\ \bibinfo {editor} {edited by\ \bibinfo {editor} {\bibfnamefont
  {H.~K.}\ \bibnamefont {Rae}}}\ (\bibinfo  {publisher} {UTC},\ \bibinfo {year}
  {1978})\ pp.\ \bibinfo {pages} {1--26}\BibitemShut {NoStop}%
\bibitem [{\citenamefont {Beenakker}, \citenamefont {Borman},\ and\
  \citenamefont {Krylov}(1994)}]{Beenakker1994}%
  \BibitemOpen
  \bibfield  {author} {\bibinfo {author} {\bibfnamefont {J.~J.~M.}\
  \bibnamefont {Beenakker}}, \bibinfo {author} {\bibfnamefont {V.~D.}\
  \bibnamefont {Borman}}, \ and\ \bibinfo {author} {\bibfnamefont {S.~Y.}\
  \bibnamefont {Krylov}},\ }\href {\doibase 10.1103/PhysRevLett.72.514}
  {\bibfield  {journal} {\bibinfo  {journal} {Physical Review Letters}\
  }\textbf {\bibinfo {volume} {72}},\ \bibinfo {pages} {514} (\bibinfo {year}
  {1994})}\BibitemShut {NoStop}%
\bibitem [{\citenamefont {Wang}\ \emph {et~al.}(1999)\citenamefont {Wang},
  \citenamefont {Challa}, \citenamefont {Sholl},\ and\ \citenamefont
  {Johnson}}]{Wang1999}%
  \BibitemOpen
  \bibfield  {author} {\bibinfo {author} {\bibfnamefont {Q.}~\bibnamefont
  {Wang}}, \bibinfo {author} {\bibfnamefont {S.}~\bibnamefont {Challa}},
  \bibinfo {author} {\bibfnamefont {D.}~\bibnamefont {Sholl}}, \ and\ \bibinfo
  {author} {\bibfnamefont {J.}~\bibnamefont {Johnson}},\ }\href {\doibase
  10.1103/PhysRevLett.82.956} {\bibfield  {journal} {\bibinfo  {journal}
  {Physical Review Letters}\ }\textbf {\bibinfo {volume} {82}},\ \bibinfo
  {pages} {956} (\bibinfo {year} {1999})}\BibitemShut {NoStop}%
\bibitem [{\citenamefont {Lu}, \citenamefont {Goldfield},\ and\ \citenamefont
  {Gray}(2006)}]{Lu2006}%
  \BibitemOpen
  \bibfield  {author} {\bibinfo {author} {\bibfnamefont {T.}~\bibnamefont
  {Lu}}, \bibinfo {author} {\bibfnamefont {E.~E.~M.}\ \bibnamefont
  {Goldfield}}, \ and\ \bibinfo {author} {\bibfnamefont {S.~S.~K.}\
  \bibnamefont {Gray}},\ }\href {\doibase 10.1021/jp0545142} {\bibfield
  {journal} {\bibinfo  {journal} {Journal of Physical Chemistry B}\ }\textbf
  {\bibinfo {volume} {110}},\ \bibinfo {pages} {1742} (\bibinfo {year}
  {2006})}\BibitemShut {NoStop}%
\bibitem [{\citenamefont {Garberoglio}(2009)}]{Garberoglio2009}%
  \BibitemOpen
  \bibfield  {author} {\bibinfo {author} {\bibfnamefont {G.}~\bibnamefont
  {Garberoglio}},\ }\href {\doibase 10.1016/j.cplett.2008.11.065} {\bibfield
  {journal} {\bibinfo  {journal} {Chemical Physics Letters}\ }\textbf {\bibinfo
  {volume} {467}},\ \bibinfo {pages} {270} (\bibinfo {year}
  {2009})}\BibitemShut {NoStop}%
\bibitem [{\citenamefont {ullah Rather}(2020)}]{Rather2020}%
  \BibitemOpen
  \bibfield  {author} {\bibinfo {author} {\bibfnamefont {S.}~\bibnamefont
  {ullah Rather}},\ }\href {\doibase 10.1016/j.ijhydene.2019.12.055} {\bibfield
   {journal} {\bibinfo  {journal} {International Journal of Hydrogen Energy}\
  }\textbf {\bibinfo {volume} {45}},\ \bibinfo {pages} {4653} (\bibinfo {year}
  {2020})}\BibitemShut {NoStop}%
\bibitem [{\citenamefont {Li}\ and\ \citenamefont {Liu}(2020)}]{Li2020}%
  \BibitemOpen
  \bibfield  {author} {\bibinfo {author} {\bibfnamefont {Y.}~\bibnamefont
  {Li}}\ and\ \bibinfo {author} {\bibfnamefont {H.}~\bibnamefont {Liu}},\
  }\href {\doibase 10.1016/j.ijhydene.2020.11.139} {\bibfield  {journal}
  {\bibinfo  {journal} {International Journal of Hydrogen Energy}\ } (\bibinfo
  {year} {2020}),\ 10.1016/j.ijhydene.2020.11.139}\BibitemShut {NoStop}%
\bibitem [{\citenamefont {De~Luca}(2004)}]{DeLuca2004}%
  \BibitemOpen
  \bibfield  {author} {\bibinfo {author} {\bibfnamefont {G.}~\bibnamefont
  {De~Luca}},\ }\href {\doibase 10.1016/j.seppur.2003.07.002} {\bibfield
  {journal} {\bibinfo  {journal} {Separation and Purification Technology}\
  }\textbf {\bibinfo {volume} {36}},\ \bibinfo {pages} {215} (\bibinfo {year}
  {2004})}\BibitemShut {NoStop}%
\bibitem [{\citenamefont {Salazar}\ \emph {et~al.}(2019)\citenamefont
  {Salazar}, \citenamefont {Badawi}, \citenamefont {Radola}, \citenamefont
  {Macaud},\ and\ \citenamefont {Simon}}]{Salazar2019}%
  \BibitemOpen
  \bibfield  {author} {\bibinfo {author} {\bibfnamefont {J.~M.}\ \bibnamefont
  {Salazar}}, \bibinfo {author} {\bibfnamefont {M.}~\bibnamefont {Badawi}},
  \bibinfo {author} {\bibfnamefont {B.}~\bibnamefont {Radola}}, \bibinfo
  {author} {\bibfnamefont {M.}~\bibnamefont {Macaud}}, \ and\ \bibinfo {author}
  {\bibfnamefont {J.~M.}\ \bibnamefont {Simon}},\ }\href {\doibase
  10.1021/acs.jpcc.9b05090} {\bibfield  {journal} {\bibinfo  {journal} {Journal
  of Physical Chemistry C}\ }\textbf {\bibinfo {volume} {123}},\ \bibinfo
  {pages} {23455} (\bibinfo {year} {2019})}\BibitemShut {NoStop}%
\bibitem [{\citenamefont {Radola}\ \emph {et~al.}(2020)\citenamefont {Radola},
  \citenamefont {Giraudet}, \citenamefont {Bezverkhyy}, \citenamefont {Simon},
  \citenamefont {Salazar}, \citenamefont {Macaud},\ and\ \citenamefont
  {Bellat}}]{Radola2020NewZeolites}%
  \BibitemOpen
  \bibfield  {author} {\bibinfo {author} {\bibfnamefont {B.}~\bibnamefont
  {Radola}}, \bibinfo {author} {\bibfnamefont {M.}~\bibnamefont {Giraudet}},
  \bibinfo {author} {\bibfnamefont {I.}~\bibnamefont {Bezverkhyy}}, \bibinfo
  {author} {\bibfnamefont {J.~M.}\ \bibnamefont {Simon}}, \bibinfo {author}
  {\bibfnamefont {M.}~\bibnamefont {Salazar}}, \bibinfo {author} {\bibfnamefont
  {M.}~\bibnamefont {Macaud}}, \ and\ \bibinfo {author} {\bibfnamefont {J.~P.}\
  \bibnamefont {Bellat}},\ }\href {\doibase 10.1039/D0CP03871G} {\bibfield
  {journal} {\bibinfo  {journal} {Physical Chemistry Chemical Physics}\ ,\
  \bibinfo {pages} {24561}} (\bibinfo {year} {2020})}\BibitemShut {NoStop}%
\bibitem [{\citenamefont {Bezverkhyy}\ \emph {et~al.}(2020)\citenamefont
  {Bezverkhyy}, \citenamefont {Giraudet}, \citenamefont {Dirand}, \citenamefont
  {Macaud},\ and\ \citenamefont {Bellat}}]{Bezverkhyy2020}%
  \BibitemOpen
  \bibfield  {author} {\bibinfo {author} {\bibfnamefont {I.}~\bibnamefont
  {Bezverkhyy}}, \bibinfo {author} {\bibfnamefont {M.}~\bibnamefont
  {Giraudet}}, \bibinfo {author} {\bibfnamefont {C.}~\bibnamefont {Dirand}},
  \bibinfo {author} {\bibfnamefont {M.}~\bibnamefont {Macaud}}, \ and\ \bibinfo
  {author} {\bibfnamefont {J.-P.}\ \bibnamefont {Bellat}},\ }\href {\doibase
  10.1021/acs.jpcc.0c06902} {\bibfield  {journal} {\bibinfo  {journal} {The
  Journal of Physical Chemistry C}\ }\textbf {\bibinfo {volume} {124}},\
  \bibinfo {pages} {24756} (\bibinfo {year} {2020})}\BibitemShut {NoStop}%
\bibitem [{\citenamefont {Oh}, \citenamefont {Hirscher},\ and\ \citenamefont
  {Separation}(2016)}]{Oh2016}%
  \BibitemOpen
  \bibfield  {author} {\bibinfo {author} {\bibfnamefont {H.}~\bibnamefont
  {Oh}}, \bibinfo {author} {\bibfnamefont {M.}~\bibnamefont {Hirscher}}, \ and\
  \bibinfo {author} {\bibfnamefont {H.~I.}\ \bibnamefont {Separation}},\ }\href
  {\doibase 10.1002/ejic.201600253} {\bibfield  {journal} {\bibinfo  {journal}
  {European Journal of Inorganic Chemistry}\ }\textbf {\bibinfo {volume}
  {2016}},\ \bibinfo {pages} {4278} (\bibinfo {year} {2016})}\BibitemShut
  {NoStop}%
\bibitem [{\citenamefont {FitzGerald}\ \emph {et~al.}(2008)\citenamefont
  {FitzGerald}, \citenamefont {Allen}, \citenamefont {Landerman}, \citenamefont
  {Hopkins}, \citenamefont {Matters}, \citenamefont {Myers},\ and\
  \citenamefont {Rowsell}}]{FitzGerald2008}%
  \BibitemOpen
  \bibfield  {author} {\bibinfo {author} {\bibfnamefont {S.~A.}\ \bibnamefont
  {FitzGerald}}, \bibinfo {author} {\bibfnamefont {K.}~\bibnamefont {Allen}},
  \bibinfo {author} {\bibfnamefont {P.}~\bibnamefont {Landerman}}, \bibinfo
  {author} {\bibfnamefont {J.}~\bibnamefont {Hopkins}}, \bibinfo {author}
  {\bibfnamefont {J.}~\bibnamefont {Matters}}, \bibinfo {author} {\bibfnamefont
  {R.}~\bibnamefont {Myers}}, \ and\ \bibinfo {author} {\bibfnamefont {J.~L.}\
  \bibnamefont {Rowsell}},\ }\href {\doibase 10.1103/PhysRevB.77.224301}
  {\bibfield  {journal} {\bibinfo  {journal} {Physical Review B - Condensed
  Matter and Materials Physics}\ }\textbf {\bibinfo {volume} {77}},\ \bibinfo
  {pages} {224301} (\bibinfo {year} {2008})}\BibitemShut {NoStop}%
\bibitem [{\citenamefont {Kim}\ \emph {et~al.}(2017)\citenamefont {Kim},
  \citenamefont {Balderas-Xicoht{\'{e}}ncatl}, \citenamefont {Zhang},
  \citenamefont {Kang}, \citenamefont {Hirscher}, \citenamefont {Oh},\ and\
  \citenamefont {Moon}}]{Kim2017}%
  \BibitemOpen
  \bibfield  {author} {\bibinfo {author} {\bibfnamefont {J.~Y.}\ \bibnamefont
  {Kim}}, \bibinfo {author} {\bibfnamefont {R.}~\bibnamefont
  {Balderas-Xicoht{\'{e}}ncatl}}, \bibinfo {author} {\bibfnamefont
  {L.}~\bibnamefont {Zhang}}, \bibinfo {author} {\bibfnamefont {S.~G.}\
  \bibnamefont {Kang}}, \bibinfo {author} {\bibfnamefont {M.}~\bibnamefont
  {Hirscher}}, \bibinfo {author} {\bibfnamefont {H.}~\bibnamefont {Oh}}, \ and\
  \bibinfo {author} {\bibfnamefont {H.~R.}\ \bibnamefont {Moon}},\ }\href
  {\doibase 10.1021/jacs.7b07925} {\bibfield  {journal} {\bibinfo  {journal}
  {Journal of the American Chemical Society}\ }\textbf {\bibinfo {volume}
  {139}},\ \bibinfo {pages} {15135} (\bibinfo {year} {2017})}\BibitemShut
  {NoStop}%
\bibitem [{\citenamefont {Fitzgerald}\ \emph {et~al.}(2018)\citenamefont
  {Fitzgerald}, \citenamefont {Shinbrough}, \citenamefont {Rigdon},
  \citenamefont {Rowsell}, \citenamefont {Kapelewski}, \citenamefont {Pang},
  \citenamefont {Lawler},\ and\ \citenamefont {Forster}}]{Fitzgerald2018}%
  \BibitemOpen
  \bibfield  {author} {\bibinfo {author} {\bibfnamefont {S.~A.}\ \bibnamefont
  {Fitzgerald}}, \bibinfo {author} {\bibfnamefont {K.}~\bibnamefont
  {Shinbrough}}, \bibinfo {author} {\bibfnamefont {K.~H.}\ \bibnamefont
  {Rigdon}}, \bibinfo {author} {\bibfnamefont {J.~L.}\ \bibnamefont {Rowsell}},
  \bibinfo {author} {\bibfnamefont {M.~T.}\ \bibnamefont {Kapelewski}},
  \bibinfo {author} {\bibfnamefont {S.~H.}\ \bibnamefont {Pang}}, \bibinfo
  {author} {\bibfnamefont {K.~V.}\ \bibnamefont {Lawler}}, \ and\ \bibinfo
  {author} {\bibfnamefont {P.~M.}\ \bibnamefont {Forster}},\ }\href {\doibase
  10.1021/acs.jpcc.7b11048} {\bibfield  {journal} {\bibinfo  {journal} {Journal
  of Physical Chemistry C}\ }\textbf {\bibinfo {volume} {122}},\ \bibinfo
  {pages} {1995} (\bibinfo {year} {2018})}\BibitemShut {NoStop}%
\bibitem [{\citenamefont {Cao}\ \emph {et~al.}(2020)\citenamefont {Cao},
  \citenamefont {Ren}, \citenamefont {Gong}, \citenamefont {Huang},
  \citenamefont {Fu}, \citenamefont {Chang}, \citenamefont {Chen},
  \citenamefont {Xiao}, \citenamefont {Liu}, \citenamefont {Yang},
  \citenamefont {Zhong}, \citenamefont {Peng},\ and\ \citenamefont
  {Zhang}}]{Cao2020}%
  \BibitemOpen
  \bibfield  {author} {\bibinfo {author} {\bibfnamefont {D.}~\bibnamefont
  {Cao}}, \bibinfo {author} {\bibfnamefont {J.}~\bibnamefont {Ren}}, \bibinfo
  {author} {\bibfnamefont {Y.}~\bibnamefont {Gong}}, \bibinfo {author}
  {\bibfnamefont {H.}~\bibnamefont {Huang}}, \bibinfo {author} {\bibfnamefont
  {X.}~\bibnamefont {Fu}}, \bibinfo {author} {\bibfnamefont {M.}~\bibnamefont
  {Chang}}, \bibinfo {author} {\bibfnamefont {X.}~\bibnamefont {Chen}},
  \bibinfo {author} {\bibfnamefont {C.}~\bibnamefont {Xiao}}, \bibinfo {author}
  {\bibfnamefont {D.}~\bibnamefont {Liu}}, \bibinfo {author} {\bibfnamefont
  {Q.}~\bibnamefont {Yang}}, \bibinfo {author} {\bibfnamefont {C.}~\bibnamefont
  {Zhong}}, \bibinfo {author} {\bibfnamefont {S.}~\bibnamefont {Peng}}, \ and\
  \bibinfo {author} {\bibfnamefont {Z.}~\bibnamefont {Zhang}},\ }\href
  {\doibase 10.1039/C9TA14254A} {\bibfield  {journal} {\bibinfo  {journal}
  {Journal of Materials Chemistry A}\ }\textbf {\bibinfo {volume} {8}},\
  \bibinfo {pages} {6319} (\bibinfo {year} {2020})}\BibitemShut {NoStop}%
\bibitem [{\citenamefont {Wang}\ \emph {et~al.}(2020)\citenamefont {Wang},
  \citenamefont {Lin}, \citenamefont {Peng}, \citenamefont {Chen},
  \citenamefont {Cheng},\ and\ \citenamefont {Zhang}}]{Wang2020}%
  \BibitemOpen
  \bibfield  {author} {\bibinfo {author} {\bibfnamefont {T.}~\bibnamefont
  {Wang}}, \bibinfo {author} {\bibfnamefont {E.}~\bibnamefont {Lin}}, \bibinfo
  {author} {\bibfnamefont {Y.-L.}\ \bibnamefont {Peng}}, \bibinfo {author}
  {\bibfnamefont {Y.}~\bibnamefont {Chen}}, \bibinfo {author} {\bibfnamefont
  {P.}~\bibnamefont {Cheng}}, \ and\ \bibinfo {author} {\bibfnamefont
  {Z.}~\bibnamefont {Zhang}},\ }\href {\doibase 10.1016/j.ccr.2020.213485}
  {\bibfield  {journal} {\bibinfo  {journal} {Coordination Chemistry Reviews}\
  }\textbf {\bibinfo {volume} {423}},\ \bibinfo {pages} {213485} (\bibinfo
  {year} {2020})}\BibitemShut {NoStop}%
\bibitem [{\citenamefont {Kim}, \citenamefont {Oh},\ and\ \citenamefont
  {Moon}(2019)}]{Kim2019}%
  \BibitemOpen
  \bibfield  {author} {\bibinfo {author} {\bibfnamefont {J.~Y.}\ \bibnamefont
  {Kim}}, \bibinfo {author} {\bibfnamefont {H.}~\bibnamefont {Oh}}, \ and\
  \bibinfo {author} {\bibfnamefont {H.~R.}\ \bibnamefont {Moon}},\ }\href
  {\doibase 10.1002/adma.201805293} {\bibfield  {journal} {\bibinfo  {journal}
  {Advanced Materials}\ }\textbf {\bibinfo {volume} {31}},\ \bibinfo {pages}
  {1805293} (\bibinfo {year} {2019})}\BibitemShut {NoStop}%
\bibitem [{\citenamefont {Liu}\ \emph {et~al.}(2019)\citenamefont {Liu},
  \citenamefont {Zhang}, \citenamefont {Little}, \citenamefont {Kapil},
  \citenamefont {Ceriotti}, \citenamefont {Yang}, \citenamefont {Ding},
  \citenamefont {Holden}, \citenamefont {Balderas-Xicoht{\'{e}}ncatl},
  \citenamefont {He}, \citenamefont {Clowes}, \citenamefont {Chong},
  \citenamefont {Sch{\"{u}}tz}, \citenamefont {Chen}, \citenamefont
  {Hirscher},\ and\ \citenamefont {Cooper}}]{Liu2019}%
  \BibitemOpen
  \bibfield  {author} {\bibinfo {author} {\bibfnamefont {M.}~\bibnamefont
  {Liu}}, \bibinfo {author} {\bibfnamefont {L.}~\bibnamefont {Zhang}}, \bibinfo
  {author} {\bibfnamefont {M.~A.}\ \bibnamefont {Little}}, \bibinfo {author}
  {\bibfnamefont {V.}~\bibnamefont {Kapil}}, \bibinfo {author} {\bibfnamefont
  {M.}~\bibnamefont {Ceriotti}}, \bibinfo {author} {\bibfnamefont
  {S.}~\bibnamefont {Yang}}, \bibinfo {author} {\bibfnamefont {L.}~\bibnamefont
  {Ding}}, \bibinfo {author} {\bibfnamefont {D.~L.}\ \bibnamefont {Holden}},
  \bibinfo {author} {\bibfnamefont {R.}~\bibnamefont
  {Balderas-Xicoht{\'{e}}ncatl}}, \bibinfo {author} {\bibfnamefont
  {D.}~\bibnamefont {He}}, \bibinfo {author} {\bibfnamefont {R.}~\bibnamefont
  {Clowes}}, \bibinfo {author} {\bibfnamefont {S.~Y.}\ \bibnamefont {Chong}},
  \bibinfo {author} {\bibfnamefont {G.}~\bibnamefont {Sch{\"{u}}tz}}, \bibinfo
  {author} {\bibfnamefont {L.}~\bibnamefont {Chen}}, \bibinfo {author}
  {\bibfnamefont {M.}~\bibnamefont {Hirscher}}, \ and\ \bibinfo {author}
  {\bibfnamefont {A.~I.}\ \bibnamefont {Cooper}},\ }\href {\doibase
  10.1126/science.aax7427} {\bibfield  {journal} {\bibinfo  {journal}
  {Science}\ }\textbf {\bibinfo {volume} {366}},\ \bibinfo {pages} {613}
  (\bibinfo {year} {2019})}\BibitemShut {NoStop}%
\bibitem [{\citenamefont {Kumar}\ and\ \citenamefont
  {Bhatia}(2005)}]{Kumar2005}%
  \BibitemOpen
  \bibfield  {author} {\bibinfo {author} {\bibfnamefont {A.~V.~A.}\
  \bibnamefont {Kumar}}\ and\ \bibinfo {author} {\bibfnamefont {S.~K.}\
  \bibnamefont {Bhatia}},\ }\href {\doibase 10.1103/PhysRevLett.95.245901}
  {\bibfield  {journal} {\bibinfo  {journal} {Physical Review Letters}\
  }\textbf {\bibinfo {volume} {95}},\ \bibinfo {pages} {1} (\bibinfo {year}
  {2005})}\BibitemShut {NoStop}%
\bibitem [{\citenamefont {Kumar}, \citenamefont {Jobic},\ and\ \citenamefont
  {Bhatia}(2006)}]{Kumar2006}%
  \BibitemOpen
  \bibfield  {author} {\bibinfo {author} {\bibfnamefont {A.~V.~A.}\
  \bibnamefont {Kumar}}, \bibinfo {author} {\bibfnamefont {H.}~\bibnamefont
  {Jobic}}, \ and\ \bibinfo {author} {\bibfnamefont {S.~K.}\ \bibnamefont
  {Bhatia}},\ }\href {\doibase 10.1021/jp063034n} {\bibfield  {journal}
  {\bibinfo  {journal} {Journal of Physical Chemistry B}\ }\textbf {\bibinfo
  {volume} {110}},\ \bibinfo {pages} {16666} (\bibinfo {year}
  {2006})}\BibitemShut {NoStop}%
\bibitem [{\citenamefont {Kumar}\ and\ \citenamefont
  {Bhatia}(2008)}]{Kumar2008}%
  \BibitemOpen
  \bibfield  {author} {\bibinfo {author} {\bibfnamefont {A.~V.~A.}\
  \bibnamefont {Kumar}}\ and\ \bibinfo {author} {\bibfnamefont {S.~K.}\
  \bibnamefont {Bhatia}},\ }\href {\doibase 10.1021/jp8015358} {\bibfield
  {journal} {\bibinfo  {journal} {Journal of Physical Chemistry C}\ }\textbf
  {\bibinfo {volume} {112}},\ \bibinfo {pages} {11421} (\bibinfo {year}
  {2008})}\BibitemShut {NoStop}%
\bibitem [{\citenamefont {Mondelo-Martell}, \citenamefont
  {Huarte-Larra{\~{n}}aga},\ and\ \citenamefont
  {Manthe}(2017)}]{Mondelo-Martell2017}%
  \BibitemOpen
  \bibfield  {author} {\bibinfo {author} {\bibfnamefont {M.}~\bibnamefont
  {Mondelo-Martell}}, \bibinfo {author} {\bibfnamefont {F.}~\bibnamefont
  {Huarte-Larra{\~{n}}aga}}, \ and\ \bibinfo {author} {\bibfnamefont
  {U.}~\bibnamefont {Manthe}},\ }\href {\doibase 10.1063/1.4995550} {\bibfield
  {journal} {\bibinfo  {journal} {Journal of Chemical Physics}\ }\textbf
  {\bibinfo {volume} {147}},\ \bibinfo {pages} {084103} (\bibinfo {year}
  {2017})}\BibitemShut {NoStop}%
\bibitem [{\citenamefont {Hankel}\ \emph {et~al.}(2011)\citenamefont {Hankel},
  \citenamefont {Zhang}, \citenamefont {Nguyen}, \citenamefont {Bhatia},
  \citenamefont {Gray},\ and\ \citenamefont {Smith}}]{Hankel2011}%
  \BibitemOpen
  \bibfield  {author} {\bibinfo {author} {\bibfnamefont {M.}~\bibnamefont
  {Hankel}}, \bibinfo {author} {\bibfnamefont {H.}~\bibnamefont {Zhang}},
  \bibinfo {author} {\bibfnamefont {T.~X.}\ \bibnamefont {Nguyen}}, \bibinfo
  {author} {\bibfnamefont {S.~K.}\ \bibnamefont {Bhatia}}, \bibinfo {author}
  {\bibfnamefont {S.~K.}\ \bibnamefont {Gray}}, \ and\ \bibinfo {author}
  {\bibfnamefont {S.~C.}\ \bibnamefont {Smith}},\ }\href {\doibase
  10.1039/c0cp02235g} {\bibfield  {journal} {\bibinfo  {journal} {Physical
  Chemistry Chemical Physics}\ }\textbf {\bibinfo {volume} {13}},\ \bibinfo
  {pages} {7834} (\bibinfo {year} {2011})}\BibitemShut {NoStop}%
\bibitem [{\citenamefont {Mondelo-Martell}\ and\ \citenamefont
  {Huarte-Larra{\~{n}}aga}(2016)}]{Mondelo-Martell2016}%
  \BibitemOpen
  \bibfield  {author} {\bibinfo {author} {\bibfnamefont {M.}~\bibnamefont
  {Mondelo-Martell}}\ and\ \bibinfo {author} {\bibfnamefont {F.}~\bibnamefont
  {Huarte-Larra{\~{n}}aga}},\ }\href {\doibase 10.1021/acs.jpca.6b00467}
  {\bibfield  {journal} {\bibinfo  {journal} {Journal of Physical Chemistry A}\
  }\textbf {\bibinfo {volume} {120}},\ \bibinfo {pages} {6501} (\bibinfo {year}
  {2016})}\BibitemShut {NoStop}%
\bibitem [{\citenamefont {Zhang}, \citenamefont {Light},\ and\ \citenamefont
  {Lee}(1999)}]{Zhang1999}%
  \BibitemOpen
  \bibfield  {author} {\bibinfo {author} {\bibfnamefont {D.~H.}\ \bibnamefont
  {Zhang}}, \bibinfo {author} {\bibfnamefont {J.~C.}\ \bibnamefont {Light}}, \
  and\ \bibinfo {author} {\bibfnamefont {S.-Y.~Y.}\ \bibnamefont {Lee}},\
  }\href {\doibase 10.1063/1.479870} {\bibfield  {journal} {\bibinfo  {journal}
  {Journal of Chemical Physics}\ }\textbf {\bibinfo {volume} {111}},\ \bibinfo
  {pages} {5741} (\bibinfo {year} {1999})}\BibitemShut {NoStop}%
\bibitem [{\citenamefont {Doll}\ and\ \citenamefont {Voter}(1987)}]{Doll1987}%
  \BibitemOpen
  \bibfield  {author} {\bibinfo {author} {\bibfnamefont {J.~D.}\ \bibnamefont
  {Doll}}\ and\ \bibinfo {author} {\bibfnamefont {A.~F.}\ \bibnamefont
  {Voter}},\ }\href {\doibase 10.1146/annurev.pc.38.100187.002213} {\bibfield
  {journal} {\bibinfo  {journal} {Annual Review of Physical Chemistry}\
  }\textbf {\bibinfo {volume} {38}},\ \bibinfo {pages} {413} (\bibinfo {year}
  {1987})}\BibitemShut {NoStop}%
\bibitem [{\citenamefont {Barth}(2000)}]{Barth2000}%
  \BibitemOpen
  \bibfield  {author} {\bibinfo {author} {\bibfnamefont {J.}~\bibnamefont
  {Barth}},\ }\href {\doibase 10.1016/S0167-5729(00)00002-9} {\bibfield
  {journal} {\bibinfo  {journal} {Surface Science Reports}\ }\textbf {\bibinfo
  {volume} {40}},\ \bibinfo {pages} {75} (\bibinfo {year} {2000})}\BibitemShut
  {NoStop}%
\bibitem [{\citenamefont {Yamamoto}(1960)}]{Yamamoto1960}%
  \BibitemOpen
  \bibfield  {author} {\bibinfo {author} {\bibfnamefont {T.}~\bibnamefont
  {Yamamoto}},\ }\href {\doibase 10.1063/1.1731099} {\bibfield  {journal}
  {\bibinfo  {journal} {Journal of Chemical Physics}\ }\textbf {\bibinfo
  {volume} {33}},\ \bibinfo {pages} {281} (\bibinfo {year} {1960})}\BibitemShut
  {NoStop}%
\bibitem [{\citenamefont {Miller}(1974)}]{Miller1974}%
  \BibitemOpen
  \bibfield  {author} {\bibinfo {author} {\bibfnamefont {W.~H.}\ \bibnamefont
  {Miller}},\ }\href {\doibase 10.1063/1.1682181} {\bibfield  {journal}
  {\bibinfo  {journal} {Journal of Chemical Physics}\ }\textbf {\bibinfo
  {volume} {61}},\ \bibinfo {pages} {1823} (\bibinfo {year}
  {1974})}\BibitemShut {NoStop}%
\bibitem [{\citenamefont {Miller}, \citenamefont {Schwartz},\ and\
  \citenamefont {Tromp}(1983)}]{Miller1983}%
  \BibitemOpen
  \bibfield  {author} {\bibinfo {author} {\bibfnamefont {W.~H.}\ \bibnamefont
  {Miller}}, \bibinfo {author} {\bibfnamefont {S.~D.}\ \bibnamefont
  {Schwartz}}, \ and\ \bibinfo {author} {\bibfnamefont {J.~W.}\ \bibnamefont
  {Tromp}},\ }\href {\doibase 10.1063/1.445581} {\bibfield  {journal} {\bibinfo
   {journal} {Journal of Chemical Physics}\ }\textbf {\bibinfo {volume} {79}},\
  \bibinfo {pages} {4889} (\bibinfo {year} {1983})}\BibitemShut {NoStop}%
\bibitem [{\citenamefont {Matzkies}\ and\ \citenamefont
  {Manthe}(1998)}]{Matzkies1998}%
  \BibitemOpen
  \bibfield  {author} {\bibinfo {author} {\bibfnamefont {F.}~\bibnamefont
  {Matzkies}}\ and\ \bibinfo {author} {\bibfnamefont {U.}~\bibnamefont
  {Manthe}},\ }\href {\doibase 10.1063/1.475892} {\bibfield  {journal}
  {\bibinfo  {journal} {Journal of Chemical Physics}\ }\textbf {\bibinfo
  {volume} {108}},\ \bibinfo {pages} {4828} (\bibinfo {year}
  {1998})}\BibitemShut {NoStop}%
\bibitem [{\citenamefont {Manthe}(2008)}]{Manthe2008a}%
  \BibitemOpen
  \bibfield  {author} {\bibinfo {author} {\bibfnamefont {U.}~\bibnamefont
  {Manthe}},\ }\href {\doibase 10.1063/1.2829404} {\bibfield  {journal}
  {\bibinfo  {journal} {Journal of Chemical Physics}\ }\textbf {\bibinfo
  {volume} {128}},\ \bibinfo {pages} {64108} (\bibinfo {year}
  {2008})}\BibitemShut {NoStop}%
\bibitem [{\citenamefont {Mondelo-Martell}\ and\ \citenamefont
  {Huarte-Larra{\~{n}}aga}(2015)}]{Mondelo-Martell2015b}%
  \BibitemOpen
  \bibfield  {author} {\bibinfo {author} {\bibfnamefont {M.}~\bibnamefont
  {Mondelo-Martell}}\ and\ \bibinfo {author} {\bibfnamefont {F.}~\bibnamefont
  {Huarte-Larra{\~{n}}aga}},\ }\href {\doibase 10.1088/1742-6596/635/3/032057}
  {\bibfield  {journal} {\bibinfo  {journal} {Journal of Physics: Conference
  Series}\ }\textbf {\bibinfo {volume} {635}},\ \bibinfo {pages} {032057}
  (\bibinfo {year} {2015})}\BibitemShut {NoStop}%
\bibitem [{\citenamefont {Garberoglio}, \citenamefont {DeKlavon},\ and\
  \citenamefont {Johnson}(2006)}]{Garberoglio2006}%
  \BibitemOpen
  \bibfield  {author} {\bibinfo {author} {\bibfnamefont {G.}~\bibnamefont
  {Garberoglio}}, \bibinfo {author} {\bibfnamefont {M.~M.}\ \bibnamefont
  {DeKlavon}}, \ and\ \bibinfo {author} {\bibfnamefont {J.~K.}\ \bibnamefont
  {Johnson}},\ }\href {\doibase 10.1021/jp054511p} {\bibfield  {journal}
  {\bibinfo  {journal} {Journal of Physical Chemistry B}\ }\textbf {\bibinfo
  {volume} {110}},\ \bibinfo {pages} {1733} (\bibinfo {year}
  {2006})}\BibitemShut {NoStop}%
\bibitem [{\citenamefont {Nguyen}, \citenamefont {Jobic},\ and\ \citenamefont
  {Bhatia}(2010)}]{Nguyen2010}%
  \BibitemOpen
  \bibfield  {author} {\bibinfo {author} {\bibfnamefont {T.~X.}\ \bibnamefont
  {Nguyen}}, \bibinfo {author} {\bibfnamefont {H.}~\bibnamefont {Jobic}}, \
  and\ \bibinfo {author} {\bibfnamefont {S.~K.}\ \bibnamefont {Bhatia}},\
  }\href {\doibase 10.1103/PhysRevLett.105.085901} {\bibfield  {journal}
  {\bibinfo  {journal} {Physical Review Letters}\ }\textbf {\bibinfo {volume}
  {105}},\ \bibinfo {pages} {085901} (\bibinfo {year} {2010})}\BibitemShut
  {NoStop}%
\bibitem [{\citenamefont {Contescu}\ \emph {et~al.}(2013)\citenamefont
  {Contescu}, \citenamefont {Zhang}, \citenamefont {Olsen}, \citenamefont
  {Mamontov}, \citenamefont {Morris},\ and\ \citenamefont
  {Gallego}}]{Contescu2013a}%
  \BibitemOpen
  \bibfield  {author} {\bibinfo {author} {\bibfnamefont {C.~I.}\ \bibnamefont
  {Contescu}}, \bibinfo {author} {\bibfnamefont {H.}~\bibnamefont {Zhang}},
  \bibinfo {author} {\bibfnamefont {R.~J.}\ \bibnamefont {Olsen}}, \bibinfo
  {author} {\bibfnamefont {E.}~\bibnamefont {Mamontov}}, \bibinfo {author}
  {\bibfnamefont {J.~R.}\ \bibnamefont {Morris}}, \ and\ \bibinfo {author}
  {\bibfnamefont {N.~C.}\ \bibnamefont {Gallego}},\ }\href {\doibase
  10.1103/PhysRevLett.110.236102} {\bibfield  {journal} {\bibinfo  {journal}
  {Physical Review Letters}\ }\textbf {\bibinfo {volume} {110}},\ \bibinfo
  {pages} {236102} (\bibinfo {year} {2013})}\BibitemShut {NoStop}%
\bibitem [{\citenamefont {Manolopoulos}(2002)}]{Manolopoulos2002}%
  \BibitemOpen
  \bibfield  {author} {\bibinfo {author} {\bibfnamefont {D.~E.}\ \bibnamefont
  {Manolopoulos}},\ }\href {\doibase 10.1063/1.1517042} {\bibfield  {journal}
  {\bibinfo  {journal} {Journal of Chemical Physics}\ }\textbf {\bibinfo
  {volume} {117}},\ \bibinfo {pages} {9552} (\bibinfo {year}
  {2002})}\BibitemShut {NoStop}%
\bibitem [{\citenamefont {Gonzalez-Lezana}, \citenamefont {Rackham},\ and\
  \citenamefont {Manolopoulos}(2004)}]{Gonzalez-Lezana2004}%
  \BibitemOpen
  \bibfield  {author} {\bibinfo {author} {\bibfnamefont {T.}~\bibnamefont
  {Gonzalez-Lezana}}, \bibinfo {author} {\bibfnamefont {E.~J.}\ \bibnamefont
  {Rackham}}, \ and\ \bibinfo {author} {\bibfnamefont {D.~E.}\ \bibnamefont
  {Manolopoulos}},\ }\href {\doibase 10.1063/1.1637584} {\bibfield  {journal}
  {\bibinfo  {journal} {Journal of Chemical Physics}\ }\textbf {\bibinfo
  {volume} {120}},\ \bibinfo {pages} {2247} (\bibinfo {year}
  {2004})}\BibitemShut {NoStop}%
\end{thebibliography}
%

\end{document}